\documentclass[aps,pra,twocolumn,showpacs,superscriptaddress]{revtex4}
\usepackage{graphicx}
\usepackage{amsmath}
\usepackage{amssymb}

\input{epsf}
\begin{document}

\title{
Breakdown of universality for
unequal-mass Fermi gases with infinite scattering length}

\author{D. Blume and K. M. Daily}
\affiliation{Department of Physics and Astronomy,
Washington State University,
  Pullman, Washington 99164-2814, USA}

\date{\today}

\begin{abstract}
We 
treat small trapped unequal-mass 
two-component Fermi gases at unitarity
within a non-perturbative microscopic framework
and
investigate
the system properties
as functions of the mass ratio
$\kappa$,
and the numbers $N_{1}$ and $N_2$
of heavy and light fermions.
While equal-mass 
Fermi gases
with infinitely large interspecies $s$-wave scattering length 
$a_s$ 
are universal,
we find that
unequal mass Fermi gases are, for sufficiently large $\kappa$
 and in the regime where Efimov physics is absent,
not universal.
In particular,
the $(N_1,N_2)=(2,1)$ and $(3,1)$ systems exhibit 
three-body (3b) and four-body (4b) resonances at $\kappa = 12.314(2)$
and $10.4(2)$, respectively,
as well as
surprisingly large finite-range (FR) effects. These findings
have profound implications for
ongoing experimental efforts and  
quantum simulation proposals
that utilize
unequal-mass atomic Fermi gases.
\end{abstract}

\pacs{03.75.Ss,05.30.Fk,34.50.-s}

\maketitle
Dilute ultracold atomic Fermi gases have 
recently attracted
a great deal of
attention from the atomic, nuclear, particle and condensed
matter communities~\cite{giorgini}.
An intriguing aspect of equal-mass two-component
Fermi gases is their universality. In the regime where
the interspecies $s$-wave scattering length $a_s$ is much larger
than the range $r_0$ of the underlying two-body (2b) potential,
the few- and many-body behavior of dilute equal-mass Fermi gases
is governed by a single microscopic parameter, namely $a_s$.
Universality has enabled comparisons between systems with
vastly different length and energy scales such as
neutron matter, nuclear matter and cold atomic gases.
Since the
interspecies atom-atom $s$-wave scattering length 
can be tuned to essentially any
value in the vicinity of a Fano-Feshbach resonance,
ultracold atomic gases have emerged as an ideal
model 
system with
which to test fundamental theories such as the BCS-BEC crossover
theory and 
to engineer and probe novel
quantum phases~\cite{bloch}.

A new degree of freedom, the mass ratio $\kappa$, 
comes into play when one considers
unequal-mass Fermi systems
such as 
mixtures of non-strange and strange quarks,
mixtures of electrons with different effective masses~\cite{suhl59},
and dual-species cold atomic gases~\cite{experimentmass,theory}.
For equal number of heavy and light fermions, the mismatch
of the Fermi surfaces 
gives rise to novel quantum phases such as 
an interior gap superfluid~\cite{liu03}.
Ultracold two-component atomic Fermi gases with unequal masses
are considered to be
prominent candidates with which to realize these unique phases.
Most
proposals 
along these lines
assume that 
unequal-mass atomic Fermi systems are 
stable and universal. While this is true
for Li-K mixtures, this Letter shows that these 
assumptions are, in certain regimes, violated.

We consider few-fermion systems with 
infinitely large interspecies $s$-wave scattering length $a_s$.
The infinitely
strongly interacting three-fermion system
consisting of two mass $m_1$ fermions and one mass $m_2$ fermion
with relative orbital angular momentum $L=1$ and parity $\Pi=-1$
supports, within the zero-range (ZR) framework,
an infinite number of 3b bound states if $\kappa \gtrsim  13.607$
($\kappa=m_1/m_2$)~\cite{efimov,esry,petrov}.
The properties of these states depend,
as do those of bosonic
$L^{\Pi}=0^+$ Efimov trimers,
 on $a_s$ and a so-called 3b
parameter.
For $8.619 \lesssim \kappa \lesssim 13.607$,
3b resonances have recently been predicted to 
be accessible~\cite{nishida}.
We focus on the regime with $\kappa \lesssim 13.607$
and find:
(i)
The $(2,1)$ system interacting through a purely 
attractive FR Gaussian interaction potential with $1/a_s=0$ first 
supports, in the 
ZR limit, a 3b bound state in free space for $\kappa = 12.314(2)$.
On resonance, the 3b bound state is, similar
to a $s$-wave dimer, infinitely large.
Away from the 3b resonance,
the behavior of gas-like states of the trapped system 
is to a good approximation universal.
(ii) 
Adding a light particle to the $(2,1)$
system does not, to within our numerical resolution,
lead to a new resonance.
The $(3,1)$ system, in contrast, exhibits
a 4b
resonance at $\kappa \approx 10.4(2)$.
On the one hand, these few-body resonances open intriguing opportunities for
studying weakly-bound few-body systems.
On the other hand, these resonances lead to losses in experiments,
thereby making the study of macroscopic unequal-mass Fermi gases
more challenging.
(iii)
We find that unequal-mass few-fermion systems 
exhibit surprisingly large FR effects.
This finding is relevant 
since a number of numerical techniques are more readily
adapted to treating FR than ZR interactions. Consequently,
a full theoretical understanding
implies understanding FR effects.
Furthermore,
since realistic atom-atom 
interactions
have a finite range, comparisons between theory and experiment
have to account for FR effects.

Our calculations are performed for a trapped
Fermi gas with $N$ atoms, $N=N_1+N_2$
($N_1$ atoms with mass $m_1$ and $N_2$ atoms with
mass $m_2$). The model Hamiltonian $H$
reads
\begin{eqnarray}
\label{eq_ham}
H= 
\sum_{j=1}^{N_1} \left( \frac{-\hbar^2}{2 m_1} \nabla_{\vec{r}_j}^2 +
\frac{1}{2}m_1 \omega^2 \vec{r}_j^2 \right) + \nonumber \\
\sum_{j=N_1+1}^{N} \left( \frac{-\hbar^2}{2 m_2} \nabla_{\vec{r}_j}^2 +
\frac{1}{2}m_2 \omega^2 \vec{r}_j^2 \right) + 
\sum_{j=1}^{N_1} \sum_{k=N_1+1}^N V_{tb}(r_{jk}),
\end{eqnarray}
where $\vec{r}_j$
denotes the position vector of the $j$th fermion measured with respect to the trap
center and $V_{tb}(r_{jk})$ 
the interspecies interaction potential
(here, $r_{jk}=|\vec{r}_j-\vec{r}_k|$).
Intraspecies interactions are,
away from an odd partial wave 2b
Fano-Feshbach resonance, 
weak and 
are neglected in Eq.~(\ref{eq_ham}).
The spherically symmetric harmonic 
confinement is characterized
by the angular 
frequency
$\omega$, 
which determines
the harmonic oscillator length $a_{\mathrm{ho}}$,
$a_{\mathrm{ho}}=\sqrt{\hbar/(2 \mu \omega)}$
with $\mu=m_1m_2/(m_1+m_2)$.
Throughout, we consider the infinite scattering length
limit, i.e., $1/a_s=0$.
Our calculations are performed for
the
FR potential 
$V_{\mathrm{fr}}$,
$V_{\mathrm{fr}}(r)=-V_0 \exp[-(r/(\sqrt{2}r_0))^2]$,
with
depth $V_0$ ($V_0>0$)
and range $r_0$  ($r_0 \ll a_{\mathrm{ho}}$).
For fixed $r_0$, we
adjust 
$V_0$ 
so
that the 2b
system in free space 
is just at the verge of supporting its first $s$-wave 
bound state.
For $N=3$, our results
for $V_{\mathrm{fr}}(r)$
are compared with those for
the ZR potential 
$V_{\mathrm{zr}}$,
$V_{\mathrm{zr}}(r)=2\pi (\hbar^2 a_s/\mu) \delta(\vec{r})
\frac{\partial }{\partial r}r$.

To determine the eigenenergies of the Hamiltonian $H$,
we separate off the center-of-mass degrees of freedom $\vec{R}_{CM}$
and
solve the Schr\"odinger equation in the relative
coordinates.
For the 
unitary system with ZR
interactions, 
the relative wave
function $\Psi_{\nu}$ separates into a hyperradial part 
$F_{\nu}(R)$ and a hyperangular part $\Phi_{\nu}(\vec{\Omega})$,
$\Psi_{\nu }(R,\vec{\Omega})=R^{-(3N-4)/2} F_{\nu }(R) 
\Phi_{\nu}(\vec{\Omega})$~\cite{castin};
here,
$R$ denotes the hyperradius, 
$\mu R^2 = \sum_{j=1}^N m_j (\vec{r}_j-\vec{R}_{CM})^2$,
and $\vec{\Omega}$ collectively denotes the remaining $3N-4$
degrees of freedom.
For $N=3$,
the eigenvalues of the hyperangular Schr\"odinger equation
can be  obtained by solving a transcendental equation
(see, e.g., Ref.~\cite{russian}),
resulting in
$V_{\nu}(R)=\frac{\hbar^2(s_{\nu}^2-1/4)}{2\mu R^2}$.
For $L^{\Pi}=1^-$, the quantity $s_0$---defined
as the 
positive root of $s_0^2$---decreases from $1.773$ to $0$ as $\kappa$ increases
from 1 to 13.607.
For $\kappa \gtrsim 13.607$, $s_0$ becomes purely imaginary
and 
Efimov physics
emerges~\cite{efimov,petrov,esry}.

In a second step,
the 
hyperradial Schr\"odinger equation
\begin{eqnarray}
\label{eq_radial}
\left( \frac{-\hbar^2}{2\mu} \frac{\partial^2}{\partial R^2}
+V_{\nu}(R) + \frac{1}{2} \mu \omega^2 R^2 \right) F_{\nu }(R)=
E_{\nu } F_{\nu }(R)
\end{eqnarray}
is solved for $F_{\nu}(R)$ and $E_{\nu}$.
\begin{figure}
\vspace*{+.1cm}
\includegraphics[angle=0,width=65mm]{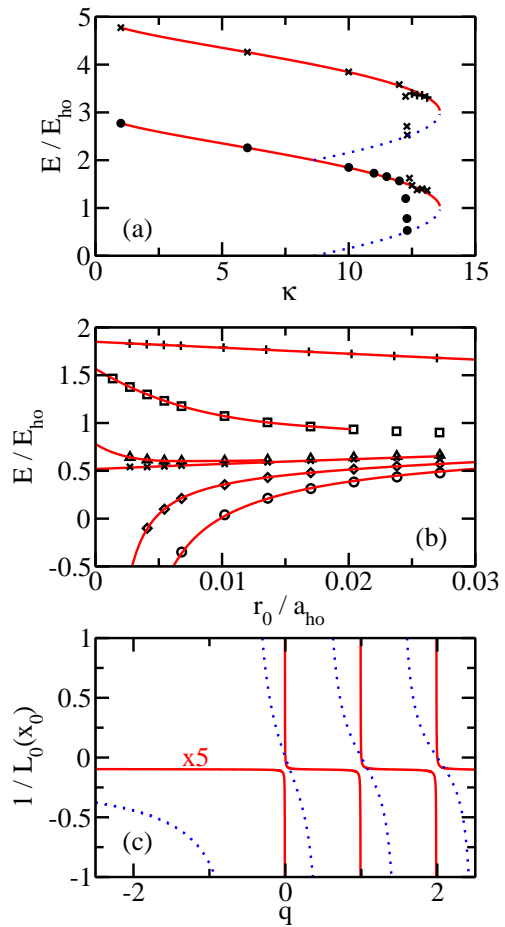}
\vspace*{0.2cm}
\caption{(Color online)
$(2,1)$ system with $L^{\Pi}=1^-$
at unitarity.
(a) 
Circles, crosses and pluses show the SV energies,
extrapolated to the ZR limit, for the three
energetically lowest-lying states  as a function of $\kappa$.
Solid and dotted lines show
$E_{f,0}$ ($q=0$ and $1$) and $E_{g,0}$ ($q=-s_0$ and $-s_0+1$),
respectively.
(b) 
Pluses, squares, triangles,
crosses, diamonds and circles show the lowest SV energy 
for $\kappa=10,12,12.3,12.314,12.4$ and $12.5$, respectively,
as a function of $r_0$.
Solid lines show fits to the SV energies.
(c)
Dotted and solid lines show $1/L_{0}(x_0)$ 
for $x_0=0.001$ as a function of $q$
for $s_0=1/2$ ($\kappa \approx  12.3131$) and $s_0=11/20$
($\kappa \approx 12.0449$), respectively. The data for $s_0=11/20$
are scaled by a factor of 5 for enhanced readibility.
In the ZR
limit,
the resonance condition implies $q=-1/2,1/2,\cdots$
for $s_0=1/2$ 
and $q=-11/20,9/20,\cdots$ for $s_0=11/20$ (see text
for details).
}\label{fig1}
\end{figure}
This
differential equation 
has two linearly independent solutions
$f_{\nu}$ and $g_{\nu}$~\cite{borca},
$f_{\nu }(x)=A_{\nu } x^{s_{\nu}+1/2} 
\exp(-x^2/2) _1 F_1(-q,s_{\nu}+1,x^2)$
and
$g_{\nu }(x)=B_{\nu } x^{-s_{\nu}+1/2} 
\exp(-x^2/2) _1 F_1(-q-s_{\nu},-s_{\nu}+1,x^2)$,
where the non-integer quantum number $q$
is defined through $E_{\nu}=(2q+s_{\nu}+1)\hbar \omega$
and $x=R/(\sqrt{2}a_{\mathrm{ho}})$.
We write 
$F_{\nu }(R)$ as
$\cos(\pi \mu_{\nu}) f_{\nu }(R)- \sin(\pi \mu_{\nu}) g_{\nu}(R)$.
Requiring
that
$F_{\nu}(R)$ vanishes at large $R$, the ``quantum defect'' 
$\mu_{\nu}$~\cite{borca}
is determined
by the condition $\sin(\pi (\mu_{\nu}+q))=0$.
Next, the allowed values of $q$ 
are
determined by
analyzing the small $R$ behavior.
The function 
$g_{\nu}$ 
is only
normalizable if $s_{\nu} < 1$~\cite{nishida,petrov};
this implies $\kappa > 8.619$
for the lowest hyperangular eigenvalue (i.e., $\nu = 0$).
Thus, for $\kappa < 8.619$, the ZR solution is determined by
the exponentially decaying piece of 
$f_{\nu}(R)$
and the 
quantization condition gives,
in agreement with Ref.~\cite{castin},
$q=0,1,\cdots$; we denote the corresponding energy by $E_{f,\nu}$
[solid lines in Fig.~\ref{fig1}(a)].
For $\kappa > 8.619$
($s_0<1$), both
$f_{\nu}$ and $g_{\nu}$ can contribute~\cite{nishida}
and the eigenenergy depends on the
boundary condition (BC) 
at small $R$. Just as in the case of Efimov trimers,
this BC
is
determined by the true atom-atom interactions
and cannot be derived within the ZR framework.
Parameterizing the BC
by the log 
derivative $L_{\nu}(x_0)$, where 
$L_{\nu}(x_0)=[F_{\nu}'(x)/F_{\nu}(x)]_{x=x_0}$,
the 
allowed $q$ values
can be obtained as a function of
$L_{\nu}(x_0)$.
Figure~\ref{fig1}(c) shows 
$[L_0(x_0)]^{-1}$ as a function of $q$ for 
$s_0=1/2$ and $11/20$.
If only $g_{\nu}$ contributes, we find,
in agreement with note [43] of Ref.~\cite{castin}, 
$q=-s_{\nu},-s_{\nu}+1,\cdots$
in the ZR limit. The corresponding energies $E_{g,0}$
are shown by dotted lines in Fig.~\ref{fig1}(a).
In the following, we determine the 3b spectrum 
for the FR potential $V_{\mathrm{fr}}(r)$ 
by the stochastic variational (SV) method,
which makes no assumption about the small $R$ behavior
of $F_{\nu}(R)$, and interpret it using Fig.~\ref{fig1}(c).

The SV approach~\cite{cgbook}
 expands the relative
wave function 
in terms of a basis set.
The 
proper anti-symmetrization of the basis functions,
which are optimized semi-stochastically, is enforced
explicitly through the application of permutation operators.
The resulting eigenenergies $E(N_1,N_2)$ of the relative
Hamiltonian
provide an upper bound to the
exact eigenenergies. 
The functional form of the basis functions used depends
on the state of interest~\cite{cgbook}.
For natural parity states, we 
use a spherical harmonic that depends on a generalized 
coordinate and multiply it by a product of Gaussians in
the relative distance coordinates.
In this case, the basis functions have definite parity
and angular momentum.
To
describe unnatural parity states,
we employ so-called geminal type basis functions that 
have neither good parity nor good angular momentum
and select the state of interest from the 
entirety of states.

Symbols in Fig.~\ref{fig1}(b) 
show selected SV
energies for the $(2,1)$ system with $L^{\Pi}=1^-$ at unitarity
as a function of $r_0$.
To extrapolate to the $r_0 \rightarrow 0$ limit,
we perform 4-5 parameter fits
to the SV energies
[solid lines in Fig.~\ref{fig1}(b)].
The resulting extrapolated ZR energies 
are shown by
symbols in Fig.~\ref{fig1}(a).
The key characteristics of 
Figs.~\ref{fig1}(a) and \ref{fig1}(b)
can be
summarized as follows:
(i) 
The dependence
of the 3b energy on $r_0$ increases 
as $\kappa$ increases from 1 to about 12. 
For $\kappa=12$, e.g.,
the difference between the extrapolated ZR energy and the 
energy for $r_0=0.01a_{\mathrm{ho}}$ is about 30\%.
For $\kappa = 12.314$, the energy depends comparatively weakly on $r_0$.
(ii) For $\kappa=12.4$ and 12.5, the ground state energy is negative
for sufficiently small $r_0$ 
and diverges as $r_0^{-2}$ with decreasing $r_0$.
(iii) For $\kappa \ne 12.314$ and $\kappa = 12.314$, 
the extrapolated ZR energies
[symbols in Fig.~\ref{fig1}(a)]
agree to a very good approximation with $E_{f,0}$ and $E_{g,0}$, 
respectively.
We interpret the dropping
of the energy around $\kappa \approx 12.3$  
($s_0 \approx 1/2$) as a 3b
resonance. The resonance position is found to be $\kappa=12.314(2)$.

Although
the occurance of the 3b resonance 
depends on the underlying 2b potential,
we now argue that a 3b resonance occurs most likely
if $s_0 \approx 1/2$.
The 
hyperradial wave function of the 3b system with FR interactions
is
for
$R \gtrsim r_0$ 
given by $F_{\nu}$. 
Figure~\ref{fig1}(c) shows the dependence of the allowed quantum numbers $q$ 
on
the value of
$1/L_0(x_0)$ for $x_0=0.001$. 
For $s_0 \ne 1/2$
[the solid line in Fig.~\ref{fig1}(c)
shows an example for $s_0=11/20$], nearly all values of the logderivative
result in $q \approx 0, 1,\cdots$,
implying 
that the realization of a 3b resonance for $s_0 \ne 1/2$ 
requires careful fine-tuning
(as $x_0$ decreases, the ``corners'' of the solid line 
near integer $q$ values become even sharper).
If realized,
such a resonance is narrow,
with the size of the zero-energy
trimer set---as in the
case of non-universal $p$-wave dimers---by the effective
angular momentum 
barrier.
Physically, an
unnaturally large contribution of $g_{\nu}$
is, for $s_0 > 1/2$, ``suppressed'' by the
effective repulsive angular momentum barrier in the hyperradial coordinate.
For $s_0=1/2$ [dotted line
in Fig.~\ref{fig1}(c)], in contrast, 
the resonance is broad and on resonance
the size of
the zero-energy trimer in free space is, much like
that of $s$-wave dimers at unitarity, infinite.
Our analysis of the structural properties such as the hyperradial
density for FR systems confirms this conclusion.
Since the effective angular momentum barrier vanishes for $s_0=1/2$,
$g_{\nu}$ is 
no longer naturally suppressed
and the existence of a 3b resonane 
is more probable. 
While our analysis shows 
a 3b resonance occurs more likely if $s_0 \approx 1/2$,
the ZR model cannot predict
whether or not such a resonance does indeed occur.
Based on the FR results presented in Figs.~\ref{fig1}(a) and (b),
we speculate that 
 other
classes of interaction potentials
likely also support a 3b resonance near $s_0 \approx 1/2$.

\begin{figure}
\vspace*{+.1cm}
\includegraphics[angle=0,width=60mm]{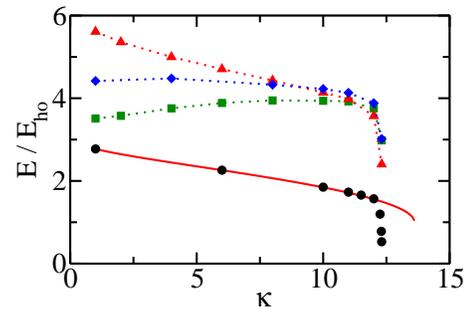}
\vspace*{0.2cm}
\caption{(Color online)
Squares, triangles and diamonds show the extrapolated 
ZR ground state energies of the $(2,2)$ 
system with $L^{\Pi}=0^{+}$, $1^{-}$ and $2^{+}$, respectively.
The extrapolation to the ZR limit 
is based on 
3-5 parameter fits
to the SV energies 
for $r_0 \approx 0.005a_{\mathrm{ho}}$ to $0.02a_{\mathrm{ho}}$. 
Circles and a solid line show the 
energies of the $(2,1)$ system with $L^{\Pi}=1^{-}$
[see Fig.~\protect\ref{fig1}(a)]. 
}\label{fig2}
\end{figure}

We now investigate the $(2,2)$ and $(3,1)$
systems, i.e., we add respectively a light and a heavy atom to the $(2,1)$
system.
Symbols in
Fig.~\ref{fig2}(a) show
the extrapolated ZR energies for the lowest natural
parity states of the $(2,2)$ system with
$L=0-2$. 
The FR effects are comparable to those of the $(2,1)$
system discussed in the context of Fig.~\ref{fig1}~\cite{vonstecher2}.
The uncertainty of the SV energies 
is 
less than 1\% and
the uncertainty of the extrapolated ZR energies is primarily due
to the fact that
our SV calculations for the $(2,2)$
systems are limited 
to $r_0 \gtrsim 0.005a_{\mathrm{ho}}$.
For all three $L$
considered, the $(2,2)$ energies
lie above the $(2,1)$
energies
and show a notable drop at $\kappa \approx 12.3$,
which we attribute to the presence of the 3b resonance.
Our calculations show no evidence for a $(2,2)$ resonance.
We also considered unnatural states of the $(2,2)$ system 
and find that the lowest unnatural parity
state lies above the lowest
natural parity state.

We now show that the $(3,1)$ system exhibits a 4b resonance at a $\kappa$ value
that differs from that at which the 3b resonance occurs.
The energetically lowest lying state of the $(3,1)$ system has 
$L=1$ and unnatural parity.
Our extrapolated ZR energies are
$5.1 \hbar \omega$, $4.9 \hbar \omega$, 
$4.5 \hbar \omega$ and $3.9 \hbar \omega$
for $\kappa=1$, 2, 4 and 8, respectively.
While our calculations for larger $\kappa$ do not
allow for a reliable extrapolation
to the ZR limit, they 
do allow for a reliable 
determination of the $(3,1)$
resonance position. To locate the 
resonance position, we 
monitor whether the SV energies
decrease or increase with decreasing $r_0$.
For $\kappa=10$, e.g., we find that the energy increases with
decreasing $r_0$.
For $\kappa=10.6$, in contrast, we find negative energies for 
$r_0 \lesssim 0.02 a_{\mathrm{ho}}$.
Performing additional calculations for $\kappa=10.3$, $10.4$
and $10.5$, 
we find that the $(3,1)$ resonance is located at $\kappa =10.4(2)$.
As in the $(2,1)$ case, the exact position of the $(3,1)$
resonance depends 
on the details of the underlying 2b interactions.
In particular, the analysis of the small $R$ BC of the
hyperradial wave function of the $(3,1)$ system parallels that of the
$(2,1)$ system. This implies that
the smallest mass ratio at which a $(3,1)$ resonance can occur 
depends on the solution to the hyperangular Schr\"odinger equation, i.e., the
$s_0$ of the $(3,1)$ system needs to be 
smaller than $1$.
It appears likely that the $(4,1)$, $(5,1)$, etc.
systems exhibit resonances at successively smaller $\kappa$.

In summary, we  considered unequal-mass two-component
Fermi gases interacting through FR potentials with infinite interspecies
$s$-wave scattering length $a_s$ and 
found 3b and 4b resonances in regimes where Efimov physics is absent.
These resonances are non-universal
in the sense that their exact position and properties depend on,
besides $a_s$, 
at least one additional parameter. 
We argued that 3b resonances occur most likely 
if $\kappa \approx 12.3$ ($s_0 \approx 1/2$)  
and that 4b systems likely exhibit, assuming a 
3b resonance exists near $s_0 \approx 1/2$,
a 4b resonance near $\kappa \approx 10.4$.
While we adjusted $s_0$
by varying $\kappa$, experimentalists could tune the $N$-body $s_0$
by utilizing an intraspecies Fano-Feshbach resonance~\cite{nishida}.
Importantly, our results for 
the $(N_1,1)$ systems apply not only to Fermi-Fermi
mixtures but also to Fermi-Bose mixtures, making 
the $^7$Li-$^{87}$Sr system ($\kappa \approx 12.4$)---thanks
to the existence of optical Sr-Sr Fano-Feshbach 
resonances~\cite{strontium}---a promising
candidate for experimental studies.
We note that the 3b resonance discussed here differs from
the resonances discussed in Ref.~\cite{russian} for positive $a_s$.
The breakdown of universality and the instability 
of unequal-mass atomic Fermi gases
discussed in this work also provides opportunities:
Our theoretical study 
paves the way for exciting investigations
of novel
few-
and many-body systems
with simultaneous 2b and 3b or 2b and 4b resonances, 
either in a dipole trap or an optical
lattice set-up.

After completion of our manuscript,
we received a 
related, independent manuscript by S. Gandolfi and J. 
Carlson~\cite{gandolfi}; see also
talk by J. Carlson on 03/11/10, 
{\em{http://www.int.washington.edu/talks/WorkShops/int\_10\_46W/}}.

Support by the NSF 
(grant PHY-0855332) 
and ARO
as well as 
fruitful 
discussions with J. Carlson, 
S. Tan
and J. von Stecher
are gratefully acknowledged.

\end{document}